\documentclass[11pt]{article}
\usepackage{color}
\usepackage{bm}
\usepackage{slashed}
\usepackage{varioref}
\usepackage{xcolor} 
\usepackage{graphicx}
\usepackage{amsmath}
\usepackage{amssymb}
\usepackage{amsfonts}
\usepackage{amsthm}
\usepackage{array}

\usepackage[
      colorlinks=true,
      linkcolor=blue,
      urlcolor=blue,
      filecolor=blue,
      citecolor=red,
      pdfstartview=FitV,
      pdftitle={},
      pdfauthor={},
      pdfsubject={},
      pdfkeywords={},
      pdfpagemode=None,
      bookmarksopen=true
]{hyperref}

\usepackage{epsfig}

\usepackage{hyperref}
\usepackage{float}
\usepackage{graphicx}
\usepackage{caption}
\usepackage{subcaption}

\def\beq{\begin{eqnarray}}
\def\eeq{\end{eqnarray}}

\def\be{\begin{equation}}
\def\ee{\end{equation}}
\def\bea{\begin{eqnarray}}
\def\eea{\end{eqnarray}}

\def\be{\begin{equation}}
\def\ee{\end{equation}}
\def\bea{\begin{eqnarray}}
\def\eea{\end{eqnarray}}

\def\cQ{\mathcal{Q}}

\def\nn{\nonumber}
\usepackage[compress]{cite}

\numberwithin{equation}{section}

\begin{document}

\begin{center}
{\LARGE \bf The Kerr-de Sitter spacetime in Bondi coordinates}

\vspace{0.3cm}
Sk Jahanur Hoque$^{a}$ and Amitabh Virmani$^{b}$

\vspace{0.3cm}

$^{a}$Institute of Theoretical Physics,
Faculty of Mathematics and Physics, \\
Charles University,  V~Hole\v{s}ovi\v{c}k\'ach 2, 180~00 Prague 8, Czech Republic 
\\
\vspace{0.3cm}

$^{b}$Chennai Mathematical Institute, \\ H1 SIPCOT IT Park, Kelambakkam, Tamil Nadu 603103, India

\vspace{0.2cm}

\texttt{jahanur.hoque@utf.mff.cuni.cz, avirmani@cmi.ac.in}
\end{center}

\begin{abstract}
 We use zero angular momentum null geodesics in the Kerr-de Sitter spacetime to transform the metric in a generalised Bondi coordinate system.   We write the metric components explicitly.
 Next, we choose the radial coordinate to be the areal coordinate and write the asymptotic metric in the Bondi-Sachs gauge. 
\end{abstract}

%
%

\tableofcontents

\section{Introduction}

In order to define total energy in a theory of gravity boundary conditions are required. Several proposals have been made for boundary conditions at the future boundary of de Sitter spacetime~\cite{Strominger:2001pn, Anninos:2012qw}. However, as emphasised in \cite{Ashtekar:2014zfa, Ashtekar:2017dlf}, this is not desirable as the future boundary conditions amount to restricting bulk dynamics. In the light of these discussions, the Bondi-Sachs formalism \cite{Bondi:1962px, Sachs:1962wk, Madler:2016xju}---which famously clarified many subtle issues in the zero 
cosmological constant case---has received much attention in recent years with non-zero cosmological constant~\cite{Chrusciel:2016oux, Poole:2018koa, Compere:2019bua, Chrusciel:2020rlz, Chrusciel:2020rlz2,  Compere:2020lrt, Compere:2020lrt2, Kolanowski:2020}. However, we do not yet have a gauge-invariant characterisation of gravitational waves in full general relativity with non-zero cosmological constant. It is expected that bringing physically interesting  solutions in Bondi coordinates can illuminate the issues involved.

The Kerr-de Sitter spacetime is a generalisation of the rotating Kerr black hole to a solution of the Einstein's equation with a positive cosmological constant $\Lambda$. It was discovered by Carter~\cite{DWDW} in 1973. The Kerr-de Sitter space has numerous properties \cite{Akcay:2010vt} and has been a continual source of new investigations and results in black hole physics.  In this paper, we bring the Kerr-de Sitter spacetime in Bondi coordinates.
Our results can be immediately adapted to the Kerr-anti de Sitter case, though we choose to keep $\Lambda >0$.

 Our analysis is based on certain null geodesics in the Kerr-de Sitter spacetime. Hackmann et al.~\cite{Hackmann:2010zz} have extensively studied  geodesics in the Boyer-Lindquist coordinates for the  Kerr-de Sitter spacetime. In this paper,  we focus on null geodesics with zero angular momentum about the axis of symmetry. We use these geodesics to generate null hypersurfaces that are the basis of our generalised Bondi coordinates. In this coordinate system, we write the metric components explicitly.

 For the Kerr solution (with zero cosmological constant) this construction was done by Fletcher and Lun \cite{FletcherLun}.  The metric was later converted into the Bondi-Sachs form by Barnich and Troessaert \cite{Barnich:2011mi}. As will become clear in the following, the analysis of  Fletcher and Lun \cite{FletcherLun} does not admit a straightforward generalisation to the non-zero $\Lambda$ cases. It is perhaps one of the reasons why the Kerr-de Sitter metric has not been written in Bondi coordinates until now.

 The key advantage of the Fletcher-Lun approach is that the metric components for the Kerr spacetime can be written using elementary functions except for one integral, which can be done in terms of an elliptic function. For the Kerr-de Sitter spacetime, we find that we have \emph{two} functions that cannot be written in terms of elementary functions. 
 
 The rest of the paper is organised as follows.
 Zero angular momentum
null geodesics are discussed in section \ref{sec:ZANG}. 
Our generalised Bondi coordinate system is presented in section \ref{sec:GBS}. 
Some intuitive understanding of our generalised Bondi coordinate system is presented in section \ref{sec:coordinates}, where we apply our coordinate transformations to flat space and de Sitter space respectively. It is curious that most of the details of our analysis are  already present in converting pure de Sitter space from Boyer-Lindquist coordinates to generalised Bondi coordinates.   
In section \ref{sec:Bondi} we choose the radial coordinate to be the areal coordinate and write the asymptotic metric in the Bondi-Sachs gauge. In this section and in appendix \ref{sec:formalism}, the consistency of our asymptotic metric within the Bondi-Sachs formalism is established. From the asymptotic metric, we read off asymptotic functions to sub-sub-sub-leading orders (4 terms in the expansion) and check the asymptotic hypersurface and evolution equations. The relevant details for the Bondi-Sachs formalism are presented in appendix \ref{sec:formalism}. 
The results presented in this paper involve much symbolic manipulations. This would not have been possible without a modern computer algebra system, in our case \texttt{Mathematica}. Five \texttt{Mathematica} are attached as ancillary files to this submission on the arXiv. An explanation on the organisation of these files is provided in appendix \ref{sec:Mathematica}.  We close with a discussion of open problems in section \ref{sec:conclusions}.

\section{Zero angular-momentum null geodesics in the Kerr-de Sitter spacetime}
\label{sec:ZANG}
The Kerr-de Sitter metric in Boyer-Lindquist coordinates $\{\bar t, \bar r, \bar \theta, \bar \phi\}$ is
\begin{align}
ds^2 =  - \frac{\bar{\Delta}_{\bar r}}{\chi^2 \bar{\rho}^2 } \left(d\bar t - a \sin^2 \bar \theta d\bar \phi\right)^2 + \frac{\bar{\rho}^2}{\bar{\Delta}_{\bar r}} d\bar r^2 
  + \frac{\bar{\Delta}_{\bar \theta} \sin^2\bar \theta }{\chi^2 \bar{\rho}^2} (a d\bar t - \bar A^2 d\bar \phi)^2 + \frac{\bar{\rho}^2}{\bar{\Delta}_{\bar \theta}} d\bar \theta^2 \,,
\end{align}
where
\begin{align}
& \bar{\Delta}_{\bar r} = \left( 1- \frac{1}{3} \Lambda {\bar r}^2\right) ({\bar r}^2 +a^2) - 2 M {\bar r}, \\
& \bar{\Delta}_{\bar \theta} = 1 + \frac{1}{3} a^2 \Lambda \cos^2 \bar{\theta}, \\
& \chi = 1 + \frac{1}{3} a^2 \Lambda, \label{def_chi} \\
& \bar{\rho}^2 = \bar{r}^2 + a ^2 \cos^2 \bar{\theta}, \\
& \bar A^2 = (\bar r^2+a^2).
\end{align}
It satisfies Einstein equation with positive cosmological constant
\be
R_{\mu\nu} - \frac{1}{2} R g_{\mu\nu} + \Lambda g_{\mu\nu} =0.
\ee

Our starting point is the geodesic equations for the Kerr-de Sitter spacetime in Boyer-Lindquist coordinates \cite{Hackmann:2010zz}\footnote{Hackmann et al.~\cite{Hackmann:2010zz} use the mostly minus sign conventions, whereas we use the mostly plus sign conventions.}.  These equations can also be easily obtained following the construction of the Kerr-de Sitter metric as explained in \cite{Frolov:2017kze}\footnote{See the explanation on the ancillary file \texttt{NullGeodesics.nb} in appendix \ref{sec:Mathematica}.}. We will  exclusively be interested in  null geodesics with zero axial angular-momentum. 
 Setting $L_z = 0$ and $\delta = 0$ in equations (14)-(17) of \cite{Hackmann:2010zz} we get the following simplified equations:
 \begin{align}
& \bar{\rho}^2 \frac{d \bar{t}}{d\lambda} = \frac{\bar \Sigma^2 E}{\bar{\Delta}_{\bar r}} , \label{t_eq} \\
& \bar{\rho}^4 \left( \frac{d \bar{r}}{d\lambda} \right)^2 = \bar B^2 E^2 , \label{r_eq} \\
& \bar{\rho}^4 \left( \frac{d \bar{\theta}}{d\lambda} \right)^2 = \bar \Omega, \label{theta_eq} \\ 
& \bar{\rho}^2 \frac{d \bar{\phi}}{d\lambda} = \chi^2 \frac{a E}{\bar{\Delta}_{\bar r} \bar{\Delta}_{\bar \theta}} \left[ 2 M \bar{r} + \frac{1}{3} \Lambda (\bar{r}^2 +a^2) \bar{\rho}^2 \right], \label{phi_eq}
\end{align}
where  $\lambda$ is an affine parameter along the null geodesic. The conserved quantities based on the $\partial_{\bar{t}}$ and $\partial_{\bar{\phi}}$ Killing vectors are,
\begin{align}
& -E =   g_{\bar{t} \, \bar{t}} \frac{d \bar{t}}{d\lambda} + g_{\bar{t} \bar \phi} \frac{d \bar{\phi}}{d\lambda}, 
& 0 = L_z =  g_{\bar \phi \bar \phi}\frac{d \bar{\phi}}{d\lambda} + g_{\bar t \bar \phi} \frac{d \bar{t}}{d\lambda}.
\end{align}
 The remaining functions appearing in equations \eqref{t_eq}--\eqref{phi_eq} are 
\begin{align}
&\bar B^2 =   (\bar{r}^2+a^2)^2 \chi^2 - \bar{\Delta}_{\bar r} \frac{K}{E^2}, & \\
&\bar \Sigma^2 =  \chi^2 \left[ (\bar{r}^2 +a^2)^2 - a^2 \sin^2 \bar{\theta} \:\frac{\bar{\Delta}_{\bar r}}{\bar{\Delta}_{\bar \theta}}   \: \right], &\\
&\bar \Omega = b^2 ( \bar{X}^2 - \sin^2\bar{\theta} ).& \label{omega}
\end{align}
Moreover, for later convenience, we have also defined,
\begin{align}
&b^2 =  a^2 E^2 \chi^2 + K (\chi - 1) \label{def_b} \\ 
& \bar{X}^2  = \frac{\chi K }{b^2}, 
\end{align}
where $K$ is the Carter's constant. Since $\bar{X}$ is related to the Carter's constant by a simple rescaling involving other constants, it is also a constant of the geodesic motion
\be
\frac{d}{ d \lambda}  \bar{X} (\bar{r}(\lambda), \bar{\theta}(\lambda))= 0.
\ee
When $\Lambda = 0$, equations \eqref{t_eq}--\eqref{phi_eq} reduce to equations (12)--(15) of Fletcher and Lun \cite{FletcherLun}. Our notation is inspired by their notation. This notation, though slightly cumbersome, will prove very convenient in the following.

A key point to note is that the right hand side of equation \eqref{phi_eq} is a function of \emph{both $\bar r$ and $\bar \theta$ only when  $\Lambda$ and $a$ are both non-zero.} If either $\Lambda$ or $a$ is zero then the right hand side of equation \eqref{phi_eq} reduces to a function  $\bar r$ alone.  As will become clear shortly, this is one of the reasons why the  Fletcher-Lun analysis does not admit a straightforward generalisation to the non-zero $\Lambda$ cases.

\section{The Kerr-de Sitter spacetime in generalised Bondi coordinates}
\label{sec:GBS}

\subsection{Preliminaries}
Our first aim is to write the Kerr-de Sitter metric in a generalised Bondi coordinate system $\{u, r, \theta, \phi\}$. This system is defined by the conditions~\cite{FletcherLun}:
\be
g_{rr} = g_{r \theta} = g_{r \phi} = 0. \label{def_GBS}
\ee
Conditions \eqref{def_GBS} are equivalent to
\be
g^{uu} = g^{u \theta} = g^{u \phi} =0.
\ee
These conditions imply that the coordinate $u$ forms null hypersurfaces, $g^{uu} = 0$, and 
that the coordinates $\theta, \phi$ are such that the lines of constant $u, \theta, \phi$ are null, i.e., $\frac{\partial}{\partial r}$ is null.

Note that conditions  \eqref{def_GBS} are preserved on replacement of $r$ by a function of \emph{all four} coordinates. This freedom can be used to match $r$ as per the requirement of the physical problem one is interested in addressing. Bondi et al \cite{Bondi:1962px} and Sachs \cite{Sachs:1962wk} famously chose to scale the radial coordinate to be the areal coordinate 
\be
g_{\theta \theta} g_{\phi \phi} - g_{\theta \phi}^2 = r^4 \sin^2 \theta. \label{det_condition}
\ee
In this section, we only impose \eqref{def_GBS} and keep the radial variable as the original radial variable $\bar{r}$. In the next section, we will choose the radial coordinate  to satisfy \eqref{det_condition} in an asymptotic expansion.

For the new coordinates, we also impose the conditions that 
\begin{align}
&\partial_u = \partial_{\bar t}, \\ 
&\partial_\phi = \partial_{\bar \phi}.
\end{align} These conditions simply ensure that we preserve the simple form of the Killing vector fields in the new coordinates.

The following coordinate transformation is necessary and sufficient to ensure that the radial coordinate does not change and the form of the Killing vector fields in the new coordinates are $ \partial_u$ and $\partial_\phi$:
\bea
\bar r &=& r , \label{transform_r} \\
\bar \theta &=& \bar \theta (r, \theta), \label{function_bar_theta} \\
\bar t &=& u + J(r, \theta), \label{function_J} \\
\bar \phi &=& \phi + L(r, \theta). \label{function_L}
\eea
The aim is to find the three functions $\bar \theta (r, \theta), J(r, \theta)$, and $L(r, \theta)$.

\subsection{The function $\bar \theta (r, \theta)$}

We require that the integral curves of the zero angular-momentum null geodesics in the new coordinates be lines of constant $\{u, \theta, \phi\}$, i.e., 
\begin{align}
& \frac{d u}{d \lambda} =  0,  \label{condition_integral_curves_1} \\
&\frac{d \theta}{d \lambda} = 0,  \\
& \frac{d \phi}{d \lambda} = 0. \label{condition_integral_curves_3}
\end{align}
Inserting conditions \eqref{condition_integral_curves_1}--\eqref{condition_integral_curves_3}  in 
\eqref{transform_r}--\eqref{function_L} and using the geodesic equations \eqref{t_eq}--\eqref{phi_eq} we get
\begin{align}
&\frac{\partial J}{\partial r } = \frac{\bar \Sigma^2}{\bar B \bar{\Delta}_{\bar r} }, &\\
&\left( \frac{\partial \bar \theta}{\partial r } \right)^2 = \frac{\bar \Omega}{\bar B^2 E^2 },& \label{theta_eq_2} \\
&\frac{\partial L }{\partial r } =  \frac{a \chi^2 }{\bar B \bar{\Delta}_{\bar r} \bar{\Delta}_{\bar \theta}} \left[ 2 M \bar{r} + \frac{1}{3} \Lambda (\bar{r}^2 +a^2) \bar{\rho}^2 \right], &
\end{align}
together with 
\be
\frac{d r }{d \lambda} = \frac{E \bar B} {\bar \rho^2}. \label{radial_affine}
\ee
Since $\bar r = r$ it is possible to remove bars from $\bar A = A,~\bar B = B$ and $\bar{\Delta}_{\bar r } = \Delta_r$.
The choice of sign in equation \eqref{radial_affine} ensures that our radial null geodesics are outgoing (as opposed to ingoing). This equation also relates the radial coordinate $r$ to the affine parameter $\lambda$. 

As also noted earlier, $\bar X$ is a constant of motion along the geodesic motion, related to the Carter's constant. In the new coordinates, since 
\be
\frac{d \theta }{d\lambda} =0  
\ee
it follows that $\bar X (\bar r(\lambda), \bar \theta (\lambda)) = X (\theta (\lambda))$, i.e., $X$ only depends on the $\theta$ value of the geodesic and nothing else. Furthermore, for $\frac{d \bar \theta }{d\lambda}$ to be well defined we must have (cf.~eqs.~\eqref{theta_eq} and  \eqref{omega})
\be 
\bar \Omega \ge 0.
\ee 
This translates into 
\be
X^2 \ge \sin^2 \bar \theta.
\ee

We can now integrate equation \eqref{theta_eq_2}. We get
\be
\int^{\bar \theta} \frac{d\bar \theta'}{\sqrt{X^2 - \sin^2 \bar \theta'}} = \pm \int^r \frac{b }{B(r') E}dr'  =: \pm \alpha_X(r)  , \label{theta_integral}
\ee
where
\be
\frac{d \alpha_X}{d r} = \frac{b}{E B(r)}.
\ee
Recall that $b$ is a positive constant that depends on $X$, cf.~\eqref{def_b}. To emphasise this $X$ dependence we put the subscript $X$ on the function $\alpha_X(r)$.   

The integral on the left hand side of equation \eqref{theta_integral} is the Legendre incomplete elliptic integral of the first kind that defines the Jacobi elliptic sine ($\text{sn}$)  function. Let us recall the definition. Let
\be
u = \int_0^\phi \frac{d\theta}{\sqrt{1 - m \sin^2 \theta}},
\ee
for $0 < m < 1$, then, 
\be
\text{sn}(u, m) = \sin \phi. 
\ee
The Jacobi elliptic sine function satisfies  \cite[page 249]{Hancock} 
\be
  \text{sn} \left(k \, u, \frac{1}{k^2}\right) = k \: \text{sn}(u, k^2).
\ee
It then follows that~\cite{FletcherLun}
\be
\sin \bar \theta =     \begin{cases}
      \text{sn} \left(\pm  X \alpha_X + H, \frac{1}{X^2} \right) & \text{for \quad $X^2 > 1$}\\
      \tanh(\pm \alpha + H) & \text{for \quad $X^2 = 1$}\\
      X  \, \text{sn} \left(\pm  \alpha_X + H,  X^2 \right) & \text{for \quad $\sin^2 \bar \theta  \le X^2 < 1$},
    \end{cases}        \label{theta_bar_fn}
\ee
where $H (\theta) $ is an arbitrary function of $\theta$; an integration constant.  For $X=1$ we simply denote $\alpha_X$  as $\alpha$.

Next, we require $\theta \to \bar \theta$ as $ r \to \infty$. This condition fixes the function $H$ as follows
\be
H =     \begin{cases}
      \text{sn}^{-1} \left(\sin \theta, \frac{1}{X^2} \right) & \text{for \quad $X^2 > 1$}\\
      \tanh^{-1}(\sin \theta) & \text{for \quad $X^2 = 1$}\\
       \text{sn}^{-1} \left( \frac{\sin \theta}{X},  X^2 \right) & \text{for \quad $\sin^2 \bar \theta  \le X^2 < 1$}.
    \end{cases}       \label{H_fn}
\ee
Finally, we require $\theta \to \frac{\pi}{2}$ as $\bar \theta \to \frac{\pi}{2}$, that is, the equator of the old coordinates also be the equator of the new coordinates. This is a desirable condition as the equator is a natural plane of symmetry of the Kerr-de Sitter spacetime. This requirement picks out $X^2 = 1$.

The case $X^2 = 1$ offers several simplifications. Firstly, it implies that the modified Carter's constant $\cQ$ is zero,
\be
\cQ := K - \chi^2 (a E - L_z)^2 = K - \chi^2 a^2 E^2 = 0.
\ee
It also  implies 
\be
b = a E \chi^{3/2}. 
\ee
Secondly, the choice $X^2 = 1$ allows us to use elementary functions in \eqref{theta_bar_fn} and \eqref{H_fn}. From now onwards we only consider transformations with $X^2 = 1$. Accordingly, we drop the subscript $X$ from $\alpha_X(r)$.  $\alpha(r)$  is chosen to be negative, monotonically increasing function that tends to zero as $r \to \infty$,
\be
\lim_{r \to \infty} \alpha (r) = 0.
\ee
$\alpha$ can be expressed in terms of the  elliptic integral of the first kind, however, we were unable to make  further simplifications using  such an expression.
When $M=0$, a closed form expression for  $\alpha$ is (for  $r > 0$),
\be
\alpha(r) =   -   \frac{1}{2}  \log
   \left(\frac{\sqrt{r^2+ a^2}+a}{\sqrt{r^2+ a^2}-
   a}\right). \label{alpha_f}
\ee

From now onwards, we only restrict ourselves to positive large $r$ regions. The Kerr-de Sitter metric is valid for $ r< 0$ as well, but the physics issues we are interested in are all related to the asymptotic nature of the Kerr-de Sitter spacetime.   Our formulae can be adapted for small $r$ or even for $ r< 0$ regions with  minor modifications, however, we will not explore those issues in this paper. For $M\neq 0$, we found it most convenient of think of $\alpha(r)$ as a power series expansion in $1/r$. It takes the form,
\be
\alpha(r) = 
    -\frac{a}{r}+\frac{a^3}{6 r^3}+\frac{ a^3 M}{4 \chi r^4} - \frac{3 a^5}{40 r^5} - \frac{a^5 M }{4 \chi r^6} +\mathcal{O}\left(\frac{1}{r^7}\right) .
\ee

Equations \eqref{theta_bar_fn} and \eqref{H_fn} give,
\be
\tanh^{-1} (\sin \bar \theta) = \tanh^{-1} (\sin  \theta) \pm \alpha.
\ee
Next, we drop the $\pm$ sign in this equation as it is equivalent to changing the sign of $a$. Thus, we finally have, 
\be
\tanh^{-1} (\sin \bar \theta) = \tanh^{-1} (\sin  \theta) + \alpha. \label{final_bar_theta}
\ee
This   relation   determines the function $\bar{\theta}(r, \theta)$.   It also gives
\begin{align}
&\sin \bar \theta = \frac{D}{C},& \\
&\cos \bar \theta = \frac{\cos \theta}{\cosh \alpha \, C}, &
 \end{align}
where
\begin{align}
& D= \tanh \alpha + \sin \theta, & \\
& C = 1 + \tanh \alpha \sin \theta. &
\end{align}

\subsection{The remaining functions}
To summarise, the transformation we have constructed so far takes the following differential form,
\begin{align}
& \bar r = r ,& \\
& \bar \theta = \bar \theta(r, \theta), \\
&d \bar t=d u  + \frac{\bar \Sigma^2}{B \Delta_r} dr  + g(r, \theta) d \theta, \label{transform_dt} \\
&d \bar \phi = d\phi + \frac{a \chi^2 }{ B {\Delta}_{r} \bar{\Delta}_{\bar \theta}} \left[ 2 M r + \frac{1}{3} \Lambda (r^2 +a^2) \bar{\rho}^2 \right] dr + h(r, \theta) d \theta. \label{transform_dphi}
 \end{align}
Functions  $g(r, \theta) $ and $h(r, \theta) $ are yet to be determined. The function $\bar{\theta}(r, \theta)$ is given via \eqref{final_bar_theta}.

It turns out that the function $g(r, \theta) $ is uniquely fixed by the condition 
\be
 g_{r\theta} = 0.
 \ee
 We find
 \be
 g(r, \theta) = a \frac{\cos \theta}{ C^2 \cosh^2 \alpha} \frac{\chi^{3/2}}{\bar \Delta_{\bar \theta}}.
 \ee
This function satisfies the integrability requirement coming from the function $J(r, \theta)$ introduced in equation \eqref{function_J}, namely, $\partial_r \partial_\theta J(r, \theta) = \partial_\theta \partial_r J(r, \theta)$. Indeed, a calculation shows that, 
 \be
 \partial_r  g (r, \theta)  = \partial_\theta \left( \frac{\bar \Sigma^2}{B \Delta_r} \right).
 \ee

From the integrability requirement for the function $L(r, \theta)$ introduced in equation \eqref{function_L}
\be
\partial_r \partial_\theta L(r, \theta) = \partial_\theta \partial_r L(r, \theta)
\ee
we have the condition
 \be
 \partial_r h(r, \theta) = \partial_\theta \left[ \frac{a \chi^2 }{ B \Delta_r \bar{\Delta}_{\bar \theta}} \left[ 2 M r + \frac{1}{3} \Lambda (r^2 +a^2) \bar{\rho}^2 \right]  \right].
 \ee
Curiously enough for $\Lambda = 0$ or for $a=0$ it is consistent to take $h(r, \theta)=0$ but it is \emph{not} possible to do so  when $\Lambda$ and $a$ are both non-zero.  
This is related to the point mentioned at the end of section \ref{sec:ZANG} that the right hand side of equation \eqref{phi_eq} is a function of \emph{both} $\bar r$ and $\bar \theta$ \emph{only when  $\Lambda$ and $a$ are both non-zero.}  

We have
 \be
 h(r, \theta) = \int^r \partial_\theta \left[ \frac{a \chi^2 }{ B(r') {\Delta}_{ r}(r') \bar{\Delta}_{\bar \theta}} \left[ 2 M r' + \frac{1}{3} \Lambda (r'{}^2 +a^2) \bar{\rho}^2 (r', \bar \theta) \right]  \right] dr'.
 \ee 
 We chose boundary conditions such that $h(r, \theta) \to 0$ as $ r \to \infty$. $h(r, \theta)$ and $\alpha(r)$ are the two functions whose explicit forms are not easy to write.  
 
 For $M=0$ it is possible to do the integral in a simple way. We find, 
\be
h(r, \theta)\Big{|}_{M=0} =  a^3 \Lambda  \sqrt{\chi} \, \frac{ \cos
   \theta}{\Theta} \left( \frac{2 A \sin \theta  - a \, (1 + \sin^2 \theta)}{3 \, (A - a \sin \theta)^2 + a^2 \, r^2 \, \Lambda \,  \cos^2\theta}\right). \label{func_h}
 \ee
where 
\be
\Theta = 1 + \frac{1}{3} \Lambda a^2 \cos^2 \theta, \label{def_Theta}
\ee 
and $A = \sqrt{r^2 + a^2}.$

For non-zero $M$, we can compute this function in an asymptotic expansion in $r$. We find,
 \bea
  h(r, \theta) &=&  h(r, \theta)\Big{|}_{M=0}^{\text{series}} - \frac{a^5 \, M \, \Lambda \, \sin 2 \theta}{12 \sqrt{\chi} \, \Theta^2 \, r^4}  + \mathcal{O}\left(\frac{1}{r^5}\right),
\eea
where what we mean by $h(r, \theta)\Big{|}_{M=0}^{\text{series}} $ is the series expansion of $h(r, \theta)\Big{|}_{M=0}$ in inverse powers of $r$. We note that for non-zero $M$, the first correction enters at order $\frac{1}{r^4}$. For our later calculations we will only need terms to $\frac{1}{r^3}$. They take the form,
\bea
  h(r, \theta)  &=&  
   \frac{2}{3} a^3 \Lambda  \sqrt{\chi} \, \frac{\sin \theta  \cos
   \theta}{\Theta^2} \frac{1}{r} + a^2 \sqrt{\chi} \, \frac{\cos \theta}{\Theta^3} \left(4 \chi -(3 + 2 \chi) \Theta + \Theta^2\right)  \, \frac{1}{r^2} \nn  \\
 &&  + \,  a^3  \sqrt{\chi} \, \frac{\sin \theta  \cos
   \theta}{\Theta^4} \left(8 \chi -(4 + 8 \chi) \Theta + (3 + \chi) \Theta^2\right)  \, \frac{1}{r^3}  + \mathcal{O}\left(\frac{1}{r^4}\right).
 \eea

 With the choices made above, the various functions simplify as follows:
\begin{align}
& A = \sqrt{r^2 + a^2}, \label{def_A} \\
& \Delta_r = \bar \Delta_{\bar r} = (r^2 +a^2 )\left(1- \frac{1}{3} \Lambda r^2\right)- 2 M r, \\
& B^2 = \chi^2 (A^4 - a^2 \Delta_r), \\
&\rho^2 = \bar \rho^2 = (r^2 + a^2) - a^2 \sin^2 \bar{\theta} = (r^2 + a^2) - a^2 \frac{D^2}{C^2} = A^2 -  a^2 \frac{D^2}{C^2}, \\
& \Delta_\theta = \bar{\Delta}_{\bar \theta} = 1 +  \frac{1}{3}a^2 \Lambda \frac{\cos^2 \theta}{C^2 \cosh^2 \alpha}, \\
&\Sigma^2 =  \chi^2 \left[ (r^2 +a^2)^2 - a^2 \frac{D^2}{C^2} \: \frac{\Delta_r}{\Delta_{\theta}}   \: \right]. &
\end{align}
Note that $\Delta_\theta$ and $\Theta$ are very different quantities.

\subsection{The final metric}

Now we are in position to list all components of the metric in generalised Bondi coordinates. They are:
\begin{align}
&
\label{grr}
 g_{rr}= 0,&\\
&g_{r\theta}= 0,&\\
&g_{r\phi}= 0,& \\
&g_{uu}= - \frac{1}{\chi^2 \rho^2}\left( \Delta_r - a^2 \Delta_\theta \frac{D^2}{C^2} \right),&\\
&g_{ru}= - \frac{\rho^2}{B},& \\
&g_{u\phi}= -\frac{a }{\rho^2 \chi^2} \left((r^2 +a^2) \Delta_\theta - \Delta_r\right) \frac{D^2}{C^2},&\\
& g_{u\theta}= -\frac{a \cos \theta}{ C^2 \cosh^2 \alpha}  \frac{1}{\rho^2}\left( \Delta_r - a^2 \Delta_\theta \frac{D^2}{C^2} \right) \frac{1}{\sqrt{\chi} \Delta_{ \theta}}  + g_{u\phi}  h(r, \theta),& \\
&g_{\phi\phi}= \frac{1}{\rho^2 \chi^2} \frac{D^2}{C^4} \left[ (r^2 +a^2)^2 \Delta_\theta C^2 - a^2 \Delta_r D^2\right].& \\
&  g_{\theta\phi}= - a^2 \cos \theta \frac{ D^2}{C^4 \cosh^2 \alpha}\frac{1}{\rho^2} \left((r^2 +a^2) \Delta_\theta - \Delta_r\right)  \frac{1}{\sqrt{\chi} \: \Delta_\theta} + g_{\phi\phi} h(r, \theta) ,&\\
&  g_{\theta\theta}= \frac{1}{C^4 \, \Delta_\theta^2 \, \rho^2 \, \cosh^4 \alpha} \left[ C^2 \Delta_\theta \rho^4 \cosh^2 \alpha - \chi a^2 \cos^2 \theta \left( \Delta_r - a^2 \Delta_\theta \frac{D^2}{C^2}\right) \right]   \nn \\ 
& \qquad + 2 g_{\theta\phi}  h(r, \theta) - g_{\phi\phi} h(r, \theta)^2,&
\label{gthth}
 \end{align}

  The above expressions are cumbersome.

Upon setting $\Lambda = 0$ these expressions reduce to Fletcher-Lun expressions~\cite{FletcherLun}.  Upon setting $a = 0$ these expressions reduce to Schwarzschild-de Sitter metric in outgoing Eddington-Finkelstein coordinates:
\be
ds^2 =  - \left(1 - \frac{2M}{r}  - \frac{\Lambda r^2}{3} \right) du^2 - 2 du dr + r^2 (d \theta^2 + \sin^2 \theta d \phi^2).
\ee

 Upon setting $M = 0$ we get de Sitter metric in an unusual form.  The function $\alpha(r)$ takes the form \eqref{alpha_f} 
and the function $h(r, \theta)$ takes the form \eqref{func_h}. Using these functions it is straightforward to verify using Mathematica\footnote{In order for Mathematica to compute things efficiently, it is better to avoid square-roots. With the choice $r = \sqrt{y^2 - a^2}$ and $\theta = \sin^{-1} x$, the RGTC program \cite{RGTC} checks in less than 2 seconds (on a 2017 MacBook Pro running Mathematica 12)  that the metric  describes de Sitter spacetime. } that the spacetime is indeed de Sitter.
We expand on the nature of our transformation in section \ref{sec:coordinates}.

In generalised Bondi coordinates, $\frac{\partial}{\partial u}$ and $\frac{\partial}{\partial \phi}$ are the two Killing vector fields, by construction.

In Boyer-Lindquist coordinates, the Kerr-de Sitter metric is invariant under simultaneous inversion of $\bar{t}, \bar{\phi}$ coordinates. In generalised Bondi coordinates, the metric is not invariant under simultaneous inversion of $u, \phi$.

\section{The nature of our generalised Bondi coordinates}
\label{sec:coordinates}
In order to understand the nature of our generalised Bondi coordinates it is useful to examine two simpler cases: flat space $M = 0, \Lambda = 0, a \neq 0$ and de Sitter space $M = 0,  \Lambda \neq 0, a \neq 0$.
\subsection{Flat space}
\begin{figure}[t!]
	\centering
	\begin{subfigure}{0.4\textwidth}
		\includegraphics[width=\textwidth]{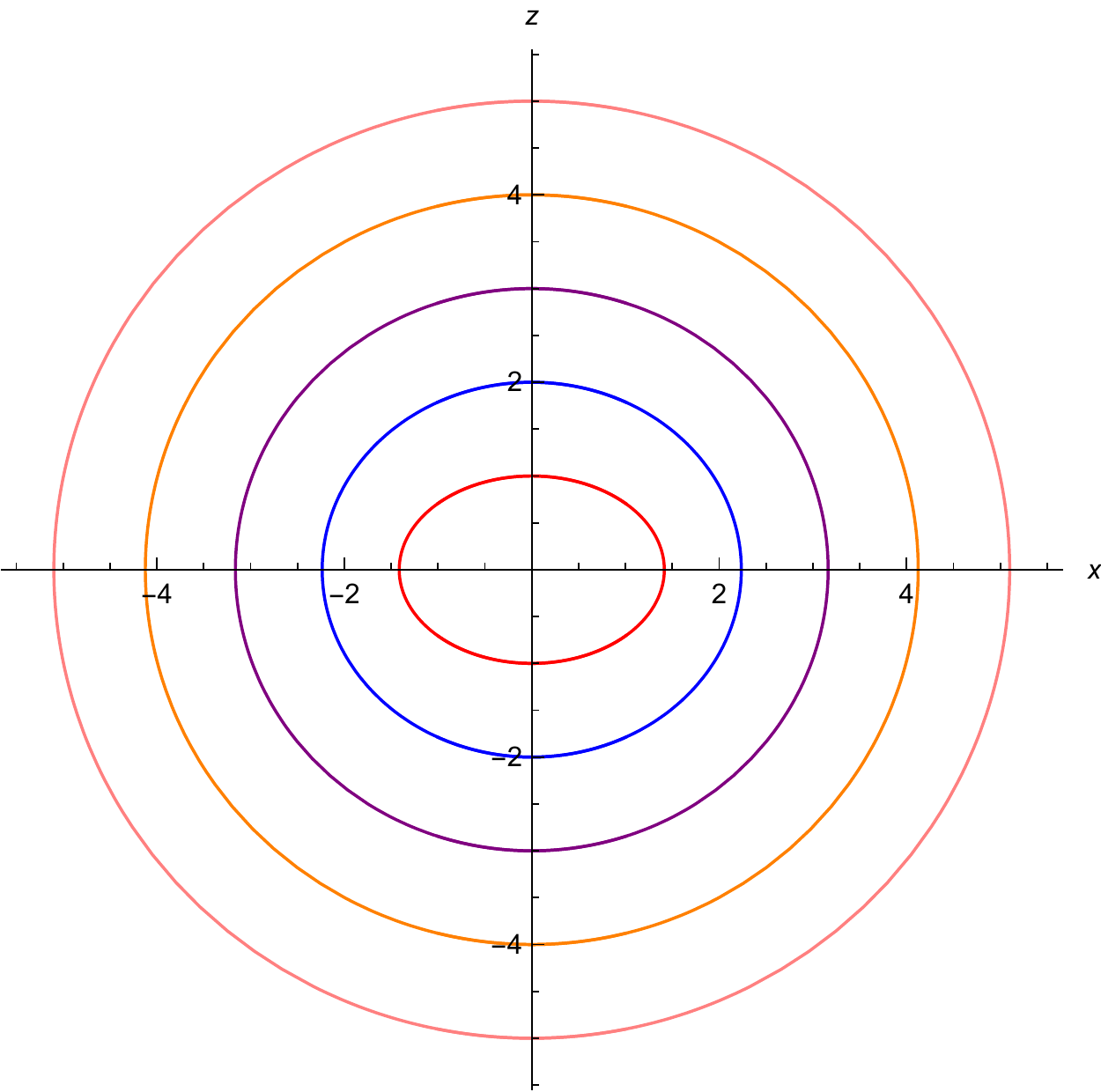}
	\end{subfigure}
	\begin{subfigure}{0.4\textwidth}
		\includegraphics[width=\textwidth]{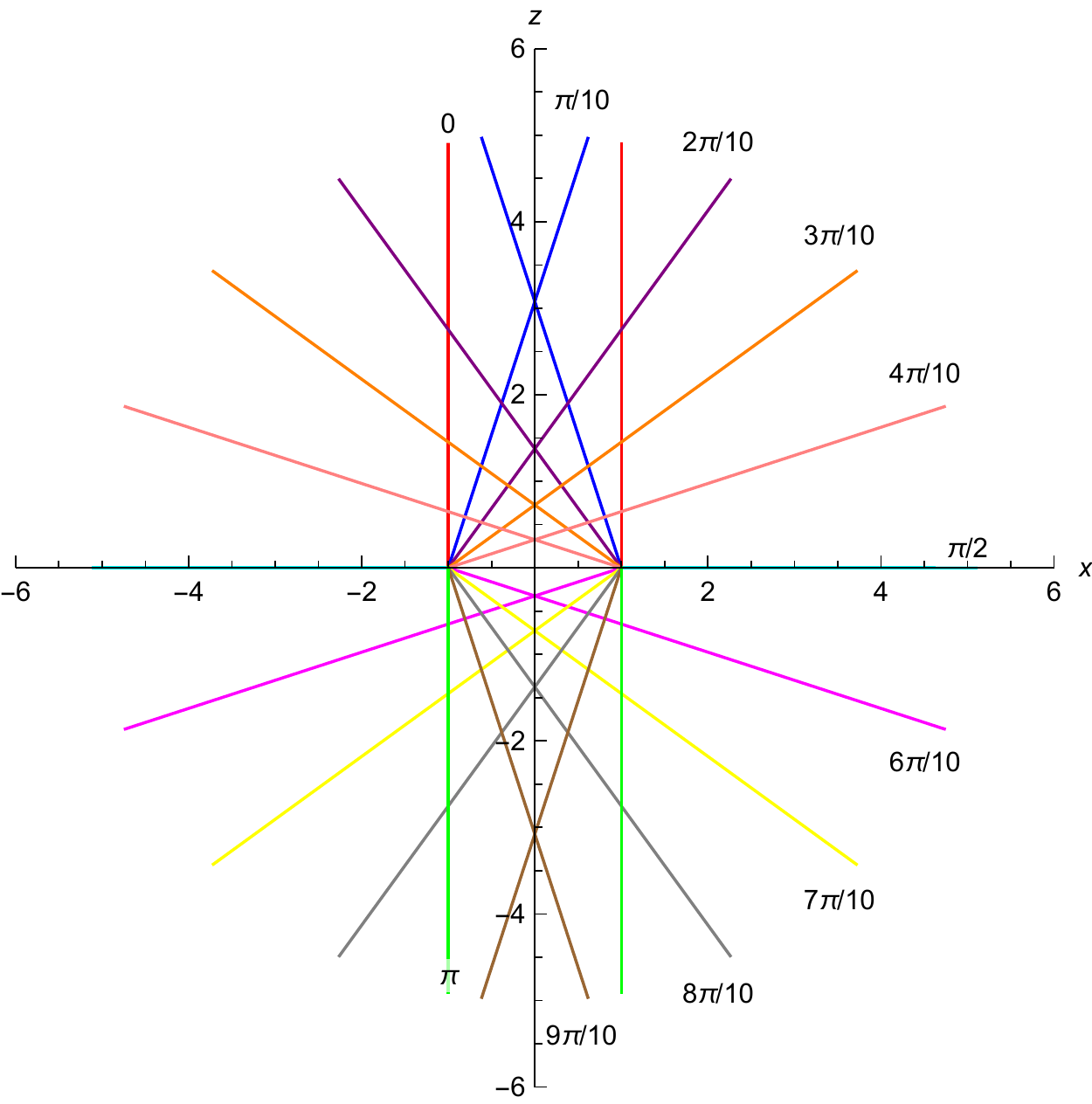}
	\end{subfigure} 
 \caption{\sf Left panel. Curves of constant $r$ with $a =1$ in the $x, z$ plane. 
The angular coordinates $\{ \theta, \phi \}$ are toroidal and the radial coordinate $r$ is confocal ellipsoidal.  Each point has a unique $r$ coordinate.   Right panel. Curves of constant $\theta$  with $a =1$ in the $x, z$ plane. Since these curves intersect, the angular coordinates are not unique. However, in the large $r$ limit $(r \gg a)$ this ambiguity disappears. Marked lines have $\phi=0$. The unmarked line of the same color has $\phi = \pi$. }
\label{fig:plot_hybrid_coordinates}

\end{figure}

Setting $M = 0, \Lambda = 0$ our metric simplifies to 
\begin{align}
\nn ds^2 =  & - du^2  - \frac{2r \left(A^2 + a^2-2 a A \sin \theta\right)}{A
   (A-a \sin \theta)^2} du dr - \frac{2 a r^2 \cos \theta}{(A-a \sin \theta)^2}du d\theta  
    +  \frac{r^4}{(A-a \sin \theta )^2} d\theta^2 \\ & + \frac{A^2 (a-A \sin \theta )^2}{(A-a \sin \theta)^2} d \phi^2, \label{flat_space}
\end{align}
where recall $A = \sqrt{r^2 + a^2}$, ~cf.~\eqref{def_A}. This metric can be constructed from the usual cartesian coordinates for flat space as follows. Starting with flat space metric 
\be
ds^2 = - dt^2 + dx^2 + dy^2 + d z^2
\ee
we introduce  toroidal coordinates~\cite{FletcherLun}
\begin{align}
&t = u + R, \\
& x = (R \sin \theta -a) \cos \phi, \\
& y = (R \sin \theta -a) \sin \phi, \\
& z = R \cos \theta.
\end{align}
\begin{figure}[t!]
\begin{center}
 \includegraphics[width=0.4\textwidth]{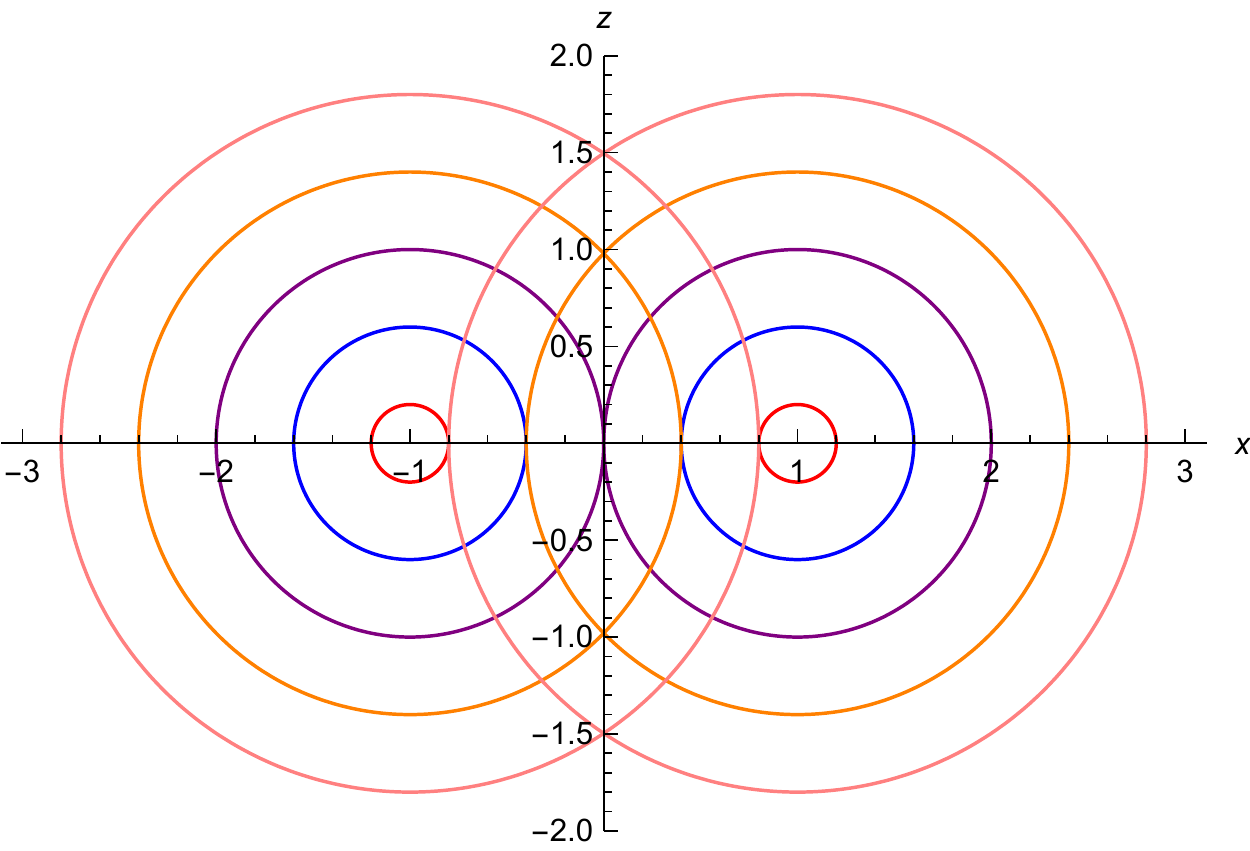}
 \caption{\sf Lines of constant $R$ with $a =1$ in the $x, z$ plane. If we restrict to $R> 0$, $-\pi <\theta \le \pi$, and $ 0 \le \phi < 2 \pi$, then each point in flat space is covered twice. The $z$-axis is covered infinitely many times. }
\label{fig:plot_toroidal_coordinates}
\end{center}
\end{figure}
The variable $R$ is such that 
\be
(\sqrt{x^2 + y^2} + a)^2 + z^2 = R^2.
\ee
If we restrict to $R> 0$, $-\pi <\theta \le \pi$, and $ 0 \le \phi < 2 \pi$, then each point in flat space is covered twice. The $z$-axis is covered infinitely many times. See figure \ref{fig:plot_toroidal_coordinates}. One problem with the double covering is that we can assign two distinct values of $R$ to a spacetime point. We can avoid this by keeping $(\theta, \phi)$ intact and by introducing the radial coordinate $r$ based on confocal ellipsoids such that 
\be
\frac{x^2 + y^2}{r^2 + a^2} + \frac{z^2}{r^2} = 1.
\ee
Explicitly 
\be
R(r, \theta) = \frac{r^2(A + a \sin \theta)}{A^2 - a^2 \sin^2 \theta}.
\ee
Now, each spacetime point has a unique $r$ and we can restrict ourselves to $0 \le \theta \le \pi$. See figure \ref{fig:plot_hybrid_coordinates}. In $\{u , r, \theta, \phi\}$ coordinates flat space metric takes the form \eqref{flat_space}.  Thus, the angular coordinates are toroidal and the radial coordinate is confocal ellipsoidal. A  discussion on these transformations can also be found in~\cite{FletcherLun}.

There is another way of thinking about the  $\{u , r, \theta, \phi\}$ that is more useful in the de Sitter space context. First, we introduce ellipsoidal coordinates (Boyer-Lindquist for flat space)
\begin{align}
&t = \bar t, \\
&x = \sqrt{{\bar r}^2 + a^2} \sin \bar \theta \cos \phi, \\
&y = \sqrt{{\bar r}^2 + a^2} \sin \bar \theta \sin \phi, \\
&z = \bar r \cos \bar \theta,
\end{align}
and then convert to $\{u, r, \theta, \phi\}$ via
\begin{align}
& \bar t  = u +  \frac{r^2}{A - a \sin \theta}, \label{flat_space_transformation_1} \\
&\bar r = r,  \label{flat_space_transformation_2} \\
&\sin \bar \theta = \frac{A \sin \theta - a}{A - a \sin \theta} ,  \label{flat_space_transformation_3} \\
&\bar \phi = \phi. \label{flat_space_transformation_4}
\end{align}
This transformation is the simplified version of the transformation \eqref{transform_r}--\eqref{function_L} via integrals of  \eqref{transform_dt} and  \eqref{transform_dphi}.

\subsection{de Sitter space}
Our starting point is the so-called `static coordinates' $\{ t_S, r_S, \theta_S, \phi_S\}$ for de Sitter space
\be
ds^2 = - \left( 1- \frac{\Lambda}{3} r_S^2\right) dt_S^2 + \left( 1- \frac{\Lambda}{3} r_S^2\right)^{-1} dr_S^2 + r_S^2 ( d\theta_S^2 + \sin^2 \theta_S d \phi_S^2).
\ee
The transformation
\begin{align}
& t_S = \frac{\bar{t}}{\chi}, & 
& r_S^2 = \frac{1}{\chi} \left( \bar {r}^2 \bar {\Delta}_{\bar \theta} + a^2 \sin^2 \bar{\theta}\right), \\
&   r_S \cos \theta_S = \bar{r} \cos \bar{\theta},& 
& \phi_S =  \bar{\phi} - \frac{a \Lambda}{3 \chi} \bar{t},
\end{align}
brings the metric in the Boyer-Lindquist form,
\begin{align}
ds^2 =  - \frac{\bar{\Delta}_{\bar r}}{\chi^2 \bar{\rho}^2 } \left(d\bar t - a \sin^2 \bar \theta d\bar \phi\right)^2 + \frac{\bar{\rho}^2}{\bar{\Delta}_{\bar r}} d\bar r^2 
  + \frac{\bar{\Delta}_{\bar \theta} \sin^2\bar \theta }{\chi^2 \bar{\rho}^2} (a d\bar t - \bar A^2 d\bar \phi)^2 + \frac{\bar{\rho}^2}{\bar{\Delta}_{\bar \theta}} d\bar \theta^2 \,,
\end{align}
where $\bar{\Delta}_{\bar r}$ now is with $M=0$, i.e., 
$\bar{\Delta}_{\bar r} =  ({\bar r}^2 +a^2) \left( 1- \frac{1}{3} \Lambda {\bar r}^2\right), 
$
and the other functions are the same as before.

The transformation \eqref{transform_r}--\eqref{function_L} to $\{u, r, \theta, \phi \}$ from  $\{\bar{t}, \bar{r}, \bar{\theta}, \bar{\phi} \}$ is  
\begin{align}
& \bar{r} = r, \\
& \sin \bar{\theta} = \frac{A \sin \theta - a}{A - a \sin \theta},  \\
& \bar t = u + J(r, \theta), \\
& \bar \phi = \phi + L(r, \theta). 
\end{align}
 with  
\begin{align}
& J(r, \theta) = \frac{\chi}{2 \lambda}  \ln \Xi, & & \Xi = \frac{ r^2 \lambda \, (a \lambda \sin \theta + \sqrt{\chi}) + A - a \sin \theta}{r^2 \lambda \, (a \lambda \sin \theta - \sqrt{\chi}) + A - a \sin \theta}, \\
& L(r, \theta) = \frac{1}{2} a \lambda \, ( \ln \Xi - \ln \Xi_\infty), & & \Xi_\infty = \lim_{r \to \infty}  \Xi =  \frac{a \lambda  \sin \theta +  \sqrt{\chi} }{a \lambda \sin \theta -  \sqrt{\chi} },
\end{align}
where we found it convenient to also introduce $ \Lambda = 3 \lambda^2$. $\lambda$ is the inverse of the dS length.

 It can be seen that this transformation reduces to flat space transformation \eqref{flat_space_transformation_1}--\eqref{flat_space_transformation_4} in the $\Lambda \to 0$ limit. The
  $\Lambda \to 0$ limit is formal for the function $L(r, \theta)$. 
 A better way to think about  this is to realise from section \ref{sec:GBS} that for $M=0$ and $\Lambda = 0$,  $L(r, \theta)=0$.

 \section{The Kerr-de Sitter metric in the Bondi-Sachs gauge}
 \label{sec:Bondi}
The Bondi-Sachs gauge is reached by defining $\widetilde{r}$ through $\det g_{AB} = \widetilde{r}^{\, 4} \sin^2 \theta$. This condition can be met order by order in an asymptotic expansion in the inverse powers of $\widetilde{r}$. It implies
 \be
r =  \chi^{1/4} \, \widetilde r +  \frac{a \cos 2 \theta}{2 \sin \theta} + \frac{a^2 \left(4 \cos 2 \theta + \csc^2 \theta \right)}{8 \,\chi^{1/4} \, \widetilde r} -   \frac{a^2 ( a \chi \cos 4 \theta + 2 M \sin \theta) }{4 \chi^{3/2} \sin \theta \, \widetilde{r}\,^2 } + \mathcal{O}\left(\frac{1}{\widetilde{r}\,^3}\right).
 \ee
 In the coordinate system $\{u, \widetilde{r}, \theta, \phi\}$,
 \be
\det g_{AB} = g_{\theta \theta} g_{\phi \phi} - g_{\theta \phi}^2 = \widetilde{r}\,{}^4 \sin^2 \theta + \mathcal{O}(\widetilde {r} \, ^0). \ee
The asymptotic metric is specified by 10 functions (for details and notation, see appendix \ref{sec:formalism})
$
\{ \beta_0, q_{AB}, U^A_0, M, N^A, E_{AB} \}
$
where
$\det q_{AB } = \sin^2 \theta$ and $q^{AB} E_{AB} =0.$
These functions for the Kerr-de Sitter metric take the form,
\begin{itemize}
\item$\beta_0$
\begin{align}
& e^{2 \beta_0} = \chi^{-5/4}
\end{align}
$\chi$ is defined in \eqref{def_chi}.
\item $q_{AB}$
\begin{align}
& q_{\theta \theta} =   \frac{ \sqrt{\chi} \left((\chi + 1) \Theta - \chi\right) }{\Theta^2}
& q_{\theta \phi} = - \frac{1}{3}a^2 \Lambda \frac{\cos \theta \sin^2 \theta}{\Theta}
\end{align}
The determinant condition fixes $q_{\phi \phi} = \chi^{-1/2} \sin^2 \theta.$ $\Theta$ is defined in \eqref{def_Theta}.

\item $U_0^A$
\begin{align}
& U_0^{\theta} = -\frac{a \Lambda \cos \theta}{3 \chi^{3/2}} & U_0^{\phi} = \frac{a \Lambda}{3 \chi \Theta}
\end{align}
\item $M(u, x^A)$
\begin{align}
&M(u, x^A)  =\frac{1}{\chi^3} \left(M - \frac{\Lambda}{6} M a^2 - \frac{a^3 \Lambda \chi}{24 \sin^3 \theta} \right)
\end{align}
\item $N^A (u, x^B)$
\begin{align}
& N^{\theta} (u, x^A)= \frac{1}{\chi^2} \left[ 3 M a \cos \theta + \frac{a^2 \chi}{8} \frac{\cos \theta}{\sin^3 \theta}\right] \\
& N^{\phi}(u, x^A) = \frac{1}{\chi^{3/2} \Theta} \left[ -3 a M + \frac{a^4 \Lambda \chi }{24} \frac{\cos^2 \theta}{\sin^3 \theta} \right]
\end{align}
\item  $E_{AB}$
\begin{align}
& E_{\theta \phi} = - \frac{a^2}{\chi^{7/4}\Theta}\left[  (1 + \chi) M \cos \theta \sin^2 \theta - \frac{\Lambda}{24} a^3 \chi \frac{\cos \theta}{\sin \theta} \right] \\
& E_{\phi \phi} = - \frac{a^2}{\chi^{9/4}}\left[ M \sin^2 \theta \, \cos 2 \theta  + \frac{a \chi}{8 \sin \theta} \right]
\end{align}
$E_{\theta \theta} $ is fixed by the trace condition $q^{AB} E_{AB} =0$. 
\end{itemize}
In appendix \ref{sec:formalism}, asymptotic equations are checked using the above expressions. 
 All equations check out perfectly. For more details, we refer the reader to \texttt{Mathematica} files.

In the limit $\Lambda \to 0$ these expressions reduce to the corresponding  expressions in \cite{Barnich:2011mi}. Note that in the Bondi-Sachs gauge some of these expressions are singular, as also in \cite{Barnich:2011mi}, at both the north and the south pole. Somewhat surprisingly the function $M(u, x^A)$---the analog of the mass-aspect---is also singular. It is not singular in the $\Lambda = 0$ limit. Unfortunately, our metric is neither in the gauge used in \cite{Compere:2019bua, Compere:2020lrt} nor in the gauge used in \cite{Chrusciel:2020rlz, Chrusciel:2020rlz2}. Therefore, we cannot immediately compute the charges using our expressions in those formalisms.

A good way to proceed is to compute the boundary stress tensor.  The pullback of the metric to the boundary takes the form
\bea
ds^2_{\mathcal{I}^+} \Big{|}_{\widetilde{r} \to \infty}^\mathrm{Bondi}&=& \Big[ \frac{\Lambda}{3}e^{4\beta_0} + U_0^A U^0_A \Big] du^2 - 2 U_A^0 du dx^A + q_{AB} dx^A dx^B \\
 &=& \frac{\Lambda}{3 \chi^{3/2}}du^2    - 2 U_A^0 du dx^A + q_{AB} dx^A dx^B.
 \label{boundary_bondi}
\eea
It is easy to check (using \texttt{Mathematica}) that the Cotton tensor of the boundary metric \eqref{boundary_bondi} vanishes, i.e., it is in the same conformal class as flat metric.

In the $r \to \infty$ limit,  transformation \eqref{transform_r}--\eqref{function_L} simplifies to,
\begin{align}
&\bar{t} = u + \frac{\chi}{2 \lambda}  \ln \frac{a \lambda  \sin \theta +  \sqrt{\chi} }{a \lambda \sin \theta -  \sqrt{\chi} }, \nn \\
&\bar{\theta} = \theta, \nn \\
&\bar{\phi} = \phi, 
\label{boundary_diff}
\end{align}
where recall $\Lambda = 3 \lambda^2$.
The boundary metric  in the Boyer-Lindquist coordinates \cite{Ashtekar:2014zfa, PremaBalakrishnan:2019jvz}
\be
ds^2_{\mathcal{I}^+} \Big{|}_{\bar{r} \to \infty}^\mathrm{BY} = \frac{1}{\chi^2}d \bar{t}^2 - \frac{2 a }{ \chi^2} \sin^2 \bar{\theta} \, d \bar{t} d \bar{\phi}  + \frac{ d \bar{\theta}^2}{\lambda^2 \bar{ \Delta}_{\bar{\theta}}} + \frac{1}{\lambda^2 \chi}\sin^2 \bar{\theta} d{\bar\phi}^2
 \label{boundary_BY}
 \ee
is related to  the boundary metric    \eqref{boundary_bondi}  in Bondi coordinates via transformation \eqref{boundary_diff} and a scaling
\be
 ds^2_{\mathcal{I}^+} \Big{|}_{\widetilde{r} \to \infty}^\mathrm{Bondi} = \omega^2 ds^2_{\mathcal{I}^+} \Big{|}_{\bar{r} \to \infty}^\mathrm{BY}
\ee
where $\omega = \lambda \chi^{1/4} $.
 The boundary stress-tensor in the Boyer-Lindquist coordinates was studied in our previous work~\cite{PremaBalakrishnan:2019jvz}. It takes the form,
\be
8 \pi G T^\mathrm{BY}_{ab} dx^a dx^b = - \frac{2 M \lambda^2 }{\chi^2} d\bar{t}^2 +  \frac{4 a \lambda^2 M \sin^2  \theta }{\chi^2}  d\bar{t} d \bar{\phi} + \frac{M}{ \bar{\Delta}_{\bar{\theta}}} d\bar{\theta}^2  +  \frac{M}{ \chi^2} \left( 3\bar{\Delta}_{\bar{\theta}} - 2 \chi \right) \sin^2 \theta  d\bar{\phi}^2.
 \label{stress_tensor_BY}
\ee
Applying \eqref{boundary_diff} we can obtain the boundary stress-tensor in Bondi coordinates,
\bea
 \label{stress_tensor_Bondi}
 T^\mathrm{Bondi}_{ab} dx^a dx^b  &=& \omega^{-1} T^\mathrm{BY}_{ab} dx^a dx^b \\
&=& T_{uu} du^2 + 2T _{u\theta} du d\theta  + 2T _{u\phi} du d\phi + T_{\theta\theta} d\theta^2 + 2 T _{\theta\phi} d\theta d\phi + T_{\phi\phi} d\phi^2
\eea
with
\begin{align}
\label{stress_tensor_Bondi_1}
&8\pi G T_{uu} = - 2 M \chi^{-9/4} \lambda  , \\
&8\pi G T _{u\phi} = 2 M a \lambda \chi^{-9/4}    \sin^2 \theta ,\\
\label{stress_tensor_Bondi_2}
&8\pi G T_{u\theta} = - 2 M a \chi^{-3/4}  \lambda \Theta^{-1} \cos \theta , \\
& 8\pi G T _{\theta\phi} = 2 M a^2 \lambda \chi^{-3/4}   \Theta^{-1} \cos \theta \sin^2 \theta,\\
& 8\pi G T_{\theta \theta} = M  \chi^{-1/4}   \lambda^{-1} \Theta^{-2} (2  \chi - (2 \chi -1) \Theta), \\  
\label{stress_tensor_Bondi_3}
&8\pi G T _{\phi\phi}= M  \chi^{-9/4} \lambda^{-1} (3 \Theta -2 \chi) \sin^2 \theta .
\end{align}

 These expressions are all regular. A short calculation shows that the charge integrals of refs.~\cite{Ashtekar:2014zfa, PremaBalakrishnan:2019jvz} using the above expressions  give the expected answers for the mass and angular momentum respectively.

This discussion clearly highlights 
that caution must be exercised in interpreting the functions $M(u, x^A)$ and $N^A(u, x^B)$.

\section{Conclusions}

\label{sec:conclusions}

Let us summarise the main findings of this paper. We  have presented the Kerr-de Sitter spacetime in a generalized Bondi coordinate system.  We wrote the metric components explicitly, using elementary functions except for two integrals $\alpha(r)$ and $h(r, \theta)$. The appearance of the function $h(r, \theta)$ is the additional complication compared to the Kerr black hole case analysis. As $\Lambda$ goes to zero this function goes to zero. Next, we have chosen the radial coordinate to be the areal coordinate and wrote the asymptotic metric in the Bondi-Sachs gauge. The consistency of our asymptotic metric within the Bondi-Sachs formalism is studied. Ref.~\cite{Compere:2019bua} had already arrived at many of the asymptotic equations we needed for our analysis. 

In  our analysis, we made certain natural choices to bring the Kerr-de Sitter spacetime in a generalized Bondi coordinate system. In these coordinates, the final metric does not satisfy the boundary gauge conditions that  \cite{Compere:2019bua} requires. Of course, it is possible to do further diffeomorphisms and bring the metric in the required form.  Though, the physics of those diffeomorphisms is not clear to us. 
We note that the relevance of the boundary gauge conditions of  \cite{Compere:2019bua} is also not understood from the linearised gravity point of view~\cite{Chrusciel:2020rlz, Chrusciel:2020rlz2, Kolanowski:2020}. See also comments in \cite{Fernandez-Alvarez:2021yog}, where a different viewpoint is advocated, and also discussion on page 13-14 of \cite{Kolanowski:2021hwo}.

In our generalised Bondi coordinates, the axis of rotational symmetry  $\sin \bar{\theta} = 0$ translates into 
\be
\sin \bar{\theta} = 0 \implies D = 0 \implies \sin \theta = - \tanh \alpha.
\ee
In the limit $r \to \infty$ the function $\alpha(r) \to 0$, so in this limit the axis of rotational symmetry is at $\sin \theta = 0$. For other values of $r$, $\alpha(r) \neq 0$. As a result, the axis of symmetry is not at $\sin \theta = 0$ but varies with the radial coordinate $r$. This is a drawback for certain numerical studies~\cite{Venter:2005cs}.  For the Kerr black hole, Pretorius and Israel \cite{Pretorius:1998sf} provided double null coordinates based on outgoing light cones. The Pretorius-Israel coordinates also serve as a useful starting point to bring the Kerr metric into the Bondi-Sachs form. The drawback here is that the coordinate transformations are implicit. The new coordinates are elliptic functions of the Boyer-Lindquist coordinates. The metric is no longer expressible in an explicit elementary form.  In \cite{Venter:2005cs} this problem was looked into. Indeed, they introduced Bondi-Sachs coordinates in which the axis of symmetry is at a fixed polar angle and the expected regularity property  is satisfied on the symmetry axis. It will be useful to generalise the construction of \cite{Venter:2005cs} to bring the Kerr-de Sitter metric into the Bondi-Sachs gauge and compare it with our study.

 It will be interesting to recover the boundary stress-tensor expressions \eqref{stress_tensor_Bondi_1}--\eqref{stress_tensor_Bondi_3} by converting the Bondi-Sachs asymptotic  form of the Kerr-de Sitter metric into the Fefferman-Graham form. This   computation can be done building upon refs.~\cite{Poole:2018koa, Compere:2019bua}, especially appendix B of \cite{Compere:2019bua}.  The details are likely to be tedious. We plan to explore this in the near future.

Despite several years to work, the subject of gravitational waves with non-zero cosmological constant is still riddled with challenges \cite{Ashtekar:2017dlf, Fernandez-Alvarez:2020}. It is expected that bringing radiative solutions in Bondi-Sachs coordinates can illuminate the issues. 
The work presented in this paper is a step in that direction. Hopefully, our  techniques can be adapted to other spacetimes, perhaps to a class of radiative spacetimes.

We hope to return to these problems in the future.

\subsection*{Acknowledgements}
The work of AV  is supported in part by the Max Planck Partnergroup ``Quantum Black Holes'' between CMI Chennai and AEI Potsdam and by a grant to CMI from the Infosys Foundation. The work of SJH is supported in part by the Czech Science Foundation Grant 19-01850S.

\appendix

 \section{The Bondi-Sachs formalism with non-zero cosmological constant}
 \label{sec:formalism}
 
Ref.~\cite{Compere:2019bua} developed certain aspects of the Bondi-Sachs formalism with non-zero cosmological constant. Some of their results can be used as a consistency check on our expressions.  In presenting our expressions in section \ref{sec:Bondi} we used their notation.
The general ansatz for the metric is
\begin{equation}
ds^2 = e^{2\beta} \frac{V}{r} du^2 - 2 e^{2\beta}du dr + g_{AB} (dx^A - U^A du)(dx^B - U^B du),
\label{bondi gauge}
\end{equation}
 where $\beta$, $U^A$, $g_{AB}$ and $V$ are arbitrary functions of the coordinates. We take the $2$-dimensional metric $g_{AB}$ to satisfy the Bondi-Sachs determinant condition \eqref{det_condition},
 \begin{equation}
 \det \, (g_{AB}) = r^4 \sin^2 \theta \label{eq:DetCond}.
\end{equation} 
 In section \ref{sec:Bondi} we used $\widetilde{r}$ to denote the areal radial coordinate; in this appendix we simply use $r$. 
Asymptotically the $2$-dimensional metric $g_{AB}$ admits an expansion
\begin{equation}
g_{AB} = r^2 \, q_{AB}  + r\, C_{AB} + D_{AB} + \frac{1}{r} \, E_{AB}  + \mathcal{O}(r^{-2})\label{eq:gABFallOff}.
\end{equation} 
Their analysis  implies
\begin{align}
&\det \, (q_{AB}) =  \sin^2 \theta \label{eq:DetCond2} , \\
&q^{AB} C_{AB} = 0, \\
&\frac{\Lambda}{3} C_{AB} = e^{-2\beta_0} \Big[ \partial_u q_{AB} + 2 D_{(A} U^0_{B)} - D^C U^0_C q_{AB} \Big],
\label{eq:CAB} \\
&D_{AB} = \frac{1}{4} q_{AB} C^{CD} C_{CD} , \\
&q^{AB} E_{AB} =0.
\end{align}
For the functions $\beta(u,r,x^A)$ and $U^A(u,r, x^B) $ their analysis gives
\begin{align}
\label{eq:EOM_beta} 
\beta(u,r,x^A) = \beta_0 (u,x^A) + \frac{1}{r^2} \Big[ -\frac{1}{32} C^{AB} C_{AB} \Big] + 
\mathcal{O}(r^{-4}).
\end{align}
and
\begin{equation}
\begin{split}
U^A = \,\, & U^A_0(u,x^B) + U^A_1(u,x^B) \frac{1}{r} +  U^A_2(u,x^B) \frac{1}{r^2} + U^A_3(u,x^B) \frac{1}{r^3}   + \mathcal{O}(r^{-4}),
\end{split} \label{eq:EOM_UA}
\end{equation}
with
\begin{eqnarray}
U^A_1(u,x^B)\hspace{-6pt} &=&\hspace{-6pt} 2 e^{2\beta_0} \partial^A \beta_0 ,\nonumber \\
U^A_2(u,x^B)\hspace{-6pt} &=&\hspace{-6pt} - e^{2\beta_0} \Big[ C^{AB} \partial_B \beta_0 + \frac{1}{2} D_B C^{AB} \Big], \nonumber\\
U^A_3(u,x^B)\hspace{-6pt} &=& \hspace{-6pt}- \frac{2}{3} e^{2\beta_0} \Big[ N^A - \frac{1}{2} C^{AB} D^C C_{BC} - \frac{3}{16} C_{CD}C^{CD} \partial^A \beta_0  \Big], \label{U3}
\end{eqnarray}
where $D_A$ is the covariant derivative defined with respect to the leading transverse metric $q_{AB}$. For the function $V(u,r,x^A)$ their analysis gives\footnote{We thank  Geoffrey Comp\`ere and Adrien Fiorucci for correspondence about their work.  Equation (2.28) in \cite{Compere:2019bua} has a minus sign typo in front of the function $M(u, x^A)$, which we have corrected in equation \eqref{V_eq}.}
\bea
\frac{V}{r} &=& \frac{\Lambda}{3} e^{2\beta_0} r^2 - r (D_A U^A_0) \label{eq:EOMVr} - e^{2\beta_0} \Big[ \frac{1}{2}\Big( R[q] + \frac{\Lambda}{8}C_{AB} C^{AB} \Big) + 2 D_A \partial^A \beta_0 + 4 \partial_A \beta_0 \partial^A \beta_0 \Big] \nonumber \\
&&  + \frac{2  M (u, x^A)}{r} + \mathcal{O}(r^{-2}). \label{V_eq}
\eea
where $R[q]$ is the two-dimensional Ricci scalar of $q_{AB}$.

Equations \eqref{U3} and \eqref{V_eq} respectively define the functions $N^A (u, x^B)$ and $M(u, x^A)$.

 We have explicitly confirmed all these equations with our asymptotic metric. In ref.~\cite{Compere:2019bua} evolution equations for $N^A (u, x^B)$ and $M(u, x^A)$ are analysed only after the boundary gauge fixing. Since our final metric is not in that gauge, those equations cannot be checked for our expressions.

\section{Explanation on \texttt{Mathematica} files}
\label{sec:Mathematica}

We have submitted five ancillary \texttt{Mathematica} files with this submission to the arXiv. A brief explanation on these files is as follows. 

\begin{enumerate}
\item \verb+NullGeodesics.nb+: In this file starting with the ``Black holes, hidden symmetries, and complete integrability'' review of 
Frolov, Krtous, and Kubiznak~\cite{Frolov:2017kze} (relevant pages, 41-51), we obtain the first order geodesic equations in the form of Hackmann, Kagramanova,  Kunz,  L\"ammerzahl~\cite{Hackmann:2010zz}. We then write these equations in our notation. 

\item \verb+KerrDeSitterBondi.nb+: This file writes the Kerr-de Sitter metric in our generalised Bondi 
coordinates. It shows that $g_{r\theta} = 0$ uniquely fixes the function $g(r, \theta)$.  Functions $\alpha(r)$ and $h(r,\theta)$ are left unspecified.

\item \verb+ExpansionFunctionH.nb+: In this file properties of the function $h(r, \theta)$ are established.

\item \verb+ExpansionKerrDeSitter.nb+: In this file,  we first compute the asymptotic metric in the generalised Bondi coordinates. Then we change the radial coordinate to the areal radial coordinate and read off asymptotic quantities. 

\item \verb+AsymptoticQuantities.nb+: In this file, we check a class of Bondi-Sachs asymptotic equations up to fourth order in the asymptotic expansion. 
\end{enumerate}

More files are available on request. 

\end{document}